\documentclass[10pt,aps,pra,showpacs,twocolumn,groupedaddress,amsmath,amssymb]{revtex4-1}

\usepackage[american]{babel}
\usepackage{graphicx}
\usepackage{times}
\usepackage{color}
\usepackage{grffile}

\newcommand{\ket}[1]{| #1 \rangle}
\newcommand{\bra}[1]{\langle #1 |}

\newcommand{\w}{{\rm{\textit w}}}
\begin{document}
\title{Complete nonclassicality test with a photon-number resolving detector}
\author{T. Kiesel and W. Vogel}


% 03.65.Ta 	Foundations of quantum mechanics; measurement theory (for optical tests of quantum theory, see 	42.50.Xa)
% 03.65.Wj 	State reconstruction, quantum tomography 
% 42.50.Dv 	Quantum state engineering and measurements 
% 42.50.Xa 	Optical tests of quantum theory 

\affiliation{Arbeitsgruppe Quantenoptik, Institut f\"ur Physik, Universit\"at  Rostock, D-18051 Rostock,
Germany}
\begin{abstract}
 We present a method for the experimental measurement of nonclassicality witnesses and demonstrate its application.  Our proposal only requires the coherent displacement of the initial state, which can be achieved by overlapping the latter with a coherent state at a beam splitter, and subsequent photon-number resolved detection. This setup allows a complete test of nonclassicality of an arbitrary quantum state. The role of the quantum efficiency as well as statistical and systematic uncertainties are discussed. Finally, the scheme is demonstrated for a realistic example.
\end{abstract}

\maketitle

\section{Introduction} 

Nonclassicality describes the difference between classical and quantum physics. For the harmonic oscillator, its formal definition is based on the coherent states $\ket\alpha$, which are seen as the closest analogues to the classical oscillation~\cite{Schrodinger}. Remarkably, any quantum state $\hat \rho$ can be formally written as a statistical mixture of (classical) coherent states~\cite{Sudarshan,Glauber},
\begin{equation}
	\hat \rho = \int d^2\alpha\,P(\alpha)\ket\alpha\bra\alpha.
\end{equation}
The function $P(\alpha)$ is the so-called Glauber Sudarshan $P$ function, playing the role of the probability density on the set of coherent states. However, it may violate the requirements for a probability density, such that it has to be considered as a quasiprobability. A state $\hat\rho$ is referred to as classical, if the $P$ function still fulfills all properties of a probability density. Otherwise, the state is called nonclassical~\cite{TitulaerGlauber}. 

However, this definition cannot be applied in most practical situations, since the $P$ function can only be understood as a generalized function, having singularities of $\delta$-type and worse. Therefore, several derived criteria for nonclassicality have been developed. 
On the one hand, one may look at different quasiprobabilities. For instance, negativities of the Wigner function are indicators for  nonclassicality of a state~\cite{Wineland,Kenfack}, but they do not appear for all nonclassical quantum states. The concept of nonclassicality filters and  quasiprobabilities is more powerful, since it provides a simple and complete characterization of nonclassical effects~\cite{Kiesel10}. First experimental demonstrations have been presented in Refs.~\cite{Kiesel11-1,Kiesel11-2,Kiesel12}. 

 On the other hand, one can examine the expectation value of a so-called nonclassicality witness $\hat W$ for a given  quantum state~\cite{Shchukin05,Korbicz05}. If this expectation value is not compatible with expectation values of classical states, then the state must be nonclassical. This criterion has been used to formulate conditions for matrices of moments~\cite{Shchukin05b} or conditions for outcome probabilities~\cite{Luis79}.

Recently, the relation between nonclassicality witnesses and quasiprobabilities has been elaborated~\cite{Kiesel-witness}, providing a unified view on all kinds of nonclassicality tests. Moreover, it has been shown that sets of nonclassicality witnesses, which enable one to detect nonclassicality of an arbitrary quantum state, can be parameterized by three real numbers only. Explicit examples for such sets of witnesses have already been given. In the present work, we show in detail how one can measure the expectation values of these witnesses, by using coherent displacement of a quantum state and subsequent photon-number resolved detection. The experimental setup is similar to the unbalanced homodyne measurement scheme~\cite{Wallentowitz96}, but adapted for our particular application. This enables us to determine the expectation values at the quantum-mechanical level of uncertainty~\cite{Kiesel-samplingnoise}, which can be used for the significant verification of nonclassical effects.

The paper is structured as follows. In Sec.~II,  we briefly review the construction of complete sets of nonclassicality witnesses. Then, we give the theoretical description of the displacement and measurement in Sec.~III, and discuss the role of statistical and systematic uncertainties. Sec.~IV is dedicated to an illustrative example. A brief summary and some conclusions are given in Sec. V.

\section{Theoretical background}

A nonclassicality witness is an operator $\hat W$, whose expectation value is nonnegative for all classical states,
\begin{equation}
	\bra\alpha\hat W\ket \alpha \geq 0.
\end{equation}
Clearly, if one examines a state $\hat \rho$ for which the expectation value of $\hat W$ is negative, then the state $\hat\rho$ is nonclassical:
\begin{equation}
	{\rm Tr}\{\hat \rho\hat W\} < 0\quad \Rightarrow \quad \hat\rho\ \mbox{nonclassical}.\label{eq_noncl_cond}
\end{equation}
 It has been shown in \cite{Kiesel-witness} that for any nonclassical state $\hat\rho$, one can verify the nonclassical character by taking a set of nonclassicality witnesses of the form
\begin{equation}
	\hat W_\w(\alpha) =  \hat D(\alpha)\hat W_\w\hat D(-\alpha),\label{eq_def_W_w(alpha)}
\end{equation}
 where $\hat D(\alpha) = e^{\alpha\hat a^\dagger-\alpha^*\hat a}$ is the coherent displacement operator, depending on the complex amplitude $\alpha$, and the operator $\hat W_\w$ is defined as
\begin{equation}
	\hat W_\w = \w^2:\!\omega^\dagger(\w \hat a^\dagger,\w\hat a)\omega(\w \hat a^\dagger,\w\hat a)\!:.
\end{equation}
Here $:\cdot:$ denotes the normal ordering procedure, and $\w$ is a real, nonnegative width parameter.  

The  function $\omega(\alpha^*,\alpha)$ can be chosen fixed. It only has to satisfy some weak properties, namely the Fourier transform of $|\omega(\alpha^*,\alpha)|^2$ shall be a nonclassicality filter fulfilling the conditions as introduced in~\cite{Kiesel10}. Roughly speaking, this means that $\omega(\alpha^*,\alpha)$ is a real continuous function, and the Fourier transform of $|\omega(\alpha^*,\alpha)|^2$ is decaying faster than any Gaussian function. A typical example of a witness operator was derived  in~\cite{Kiesel-witness},
\begin{equation}
	\hat W_\w(\alpha) = :\frac{[J_1(\w \sqrt{(\hat a^\dagger-\alpha^*)(\hat a-\alpha)})]^2}{4(\hat a^\dagger-\alpha^*)(\hat a-\alpha)}: ,
\label{eq:triangular:filter}
\end{equation} 
where $J_1(x)$ is the Bessel function of first order. Hence, a universal witness 
applying to any quantum state can even be given in a closed analytical form.
The price we have to pay is that such a witness is based on a nonclassicality filter with compact support. This means that in general it does not test the complete properties of the quantum state under study for a chosen $\w$~parameter. However, by increasing the value of $\w$, that test converges to a complete one.

To overcome this problem, we can choose a witness based on autocorrelation filters~\cite{Kiesel10,Kiesel-witness}. It tests the full state for any choice of $\w$, but can be represented only via a numerical integration. This, however, is not a serious problem since one can store the result of the numerical integration to further apply it for the direct handling of the experimental data.

\section{Measurement of nonclassicality witnesses}
\subsection{Experimental setup}
In the following, we restrict ourselves to a function of the form $\omega(\hat a^\dagger,\hat a) \equiv \omega(\hat n)$, where $\hat n=\hat a^\dagger\hat a$ is the photon number operator. In this case, the set of witnesses~\eqref{eq_def_W_w(alpha)} can be written in the form
\begin{equation}
	\hat W_\w(\alpha) = \w^2:\! \omega^\dagger(\w^2\hat n(\alpha))\omega(\w^2\hat n(\alpha))\!:,
\end{equation}
with $\hat n(\alpha) = \hat D(\alpha)\hat n\hat D(-\alpha)$ being the displaced photon number operator. In~\cite{Wallentowitz96}, a procedure has been proposed to measure the statistics of such operators. 

\begin{figure}[h]\centering
  \includegraphics[width=0.9\columnwidth]{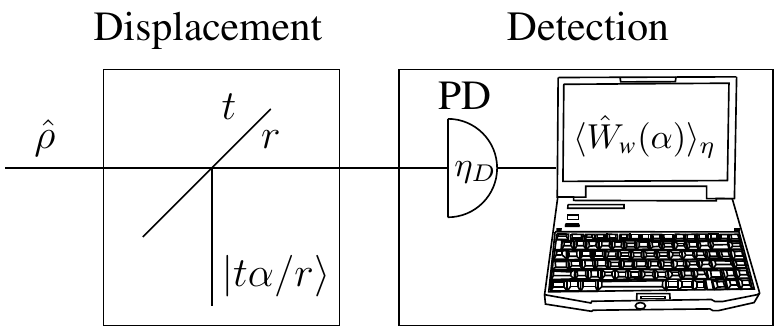}
  \caption{Experimental scheme for the detection of a nonclassicality witness. The beam splitter displaces the input state $\hat\rho$, the photodiode PD records the photon number statistics with quantum efficiency $\eta_D$. The expectation value of the witness $\hat W_\w(\alpha)$ can be calculated from Eq.~(\ref{eq:reconstruct:P}). }
  \label{fig:setup}
\end{figure}
Let us look at the sketch in Fig.~\ref{fig:setup}. The state of interest is overlapped at a beamsplitter with a coherent state of amplitude $t\alpha/r$, where $t,r$ are the transmissivity and reflectivity of the beamsplitter. The output state is measured by a photon number resolving detector with quantum efficiency $\eta_D$. It has been shown that the recorded statistics corresponds to the statistics of the operator $\hat n(\alpha)$ of the input state $\hat \rho$, measured with an effective quantum efficiency $\eta = t^2\eta_D$. Therefore, with this setup, we obtain probabilities $p_n(\alpha, \eta)$ for observing the $n$-th outcome of the measurement of the photon number operator $\hat n(\alpha)$. 

Now, we can use the witness operator $\hat W_\w$ in its spectral decomposition into photon number states $\ket n$,
\begin{equation}
	\hat W_\w = \sum_{n=0}^\infty \Omega_{\w,n} \ket n\bra n,
\end{equation}
with the matrix elements 
\begin{equation}
	\Omega_{\w,n} = \bra n :\!\omega^\dagger(\w^2\hat n)\omega(\w^2\hat n)\!:\ket n.\label{eq:matrix:elements}
\end{equation}
Note that these numbers are independent of the coherent displacement $\alpha$. All off-diagonal matrix elements are equal to zero, since we assumed that $\hat W_\w$ does only depend on the photon number operator. Analogously, we can find the spectral decomposition of the displaced witness as
\begin{equation}
	\hat W_\w(\alpha) = \sum_{n=0}^\infty \Omega_{\w,n} \ket{n(\alpha)}\bra{n(\alpha)},\label{eq_W_w(alpha)_spectral}
\end{equation}
where $\ket{n(\alpha)} = \hat D(\alpha) \ket n$ denote the eigenstates of the displaced photon number operator. Now let us calculate the expectation value of the operator in Eq.~\eqref{eq_W_w(alpha)_spectral} with respect to an input state $\hat \rho_\eta$, which suffered detection losses $\eta$. The quantity is given by
\begin{equation}
	\langle\hat W_\w(\alpha)\rangle_\eta = {\rm Tr}\{\hat \rho_\eta \hat W_\w(\alpha)\} = \sum_{n=0}^\infty p_n(\alpha,\eta) \Omega_{\w,n},\label{eq:reconstruct:P}
\end{equation}
where $p_n(\alpha,\eta) = \bra{n(\alpha)}\hat \rho_\eta\ket{n(\alpha)}$ is exactly the statistics of the photon number which is measured in the described scheme. Together with the theoretically known coefficients $\Omega_{\w,n}$, this equation allows us to estimate the expectation value of the witness.

Equation~\eqref{eq:reconstruct:P} is a central result of this manuscript, which enables us to look for negative expectation values of any witness $\hat W_\w(\alpha)$. We emphasize that for verification of nonclassicality, it is sufficient to observe negativity of the expectation value for a single complex amplitude $\alpha$. In this sense, nonclassicality can be tested locally in phase space. Conversely, to demonstrate classicality of a state, one has to examine the witnesses of all complex amplitudes $\alpha$. 

A strong point of our method consist in the following. In typical experimental scenarios one has
at least a clear conjecture about the state one is going to produce. Hence on
can easily chose the amplitude $\alpha$ to be optimal for the state one expects 
to be created. When this strategy is successful, the test only requires the optimization of one real parameter $\w$. A failure of the test proves the failure of the conjecture. In such cases one has still the option to vary the amplitude $\alpha$. 
In any case, the test procedure only requires photon-number resolved measurements in order to estimate the witness for an arbitrary amplitude $\alpha$.  We note that our witness test can be replaced by any other nonclassicality criterion, which is necessary and sufficient for photon number distributions. However, they can be much more elaborate than the calculation of a simple expectation value~\cite{NecSufPhotonNumber}.

One can minimize the losses introduced by the beamsplitter by choosing a transmissivity close to one. However, theoretically the quantum efficiency is not important for the examination of nonclassicality, since it does not destroy nonclassical effects, but may only decrease the significance of their detection. More precisely, due to the equivalence of nonclassicality witnesses and quasiprobabilities~\cite{Kiesel-witness}, we can adapt the considerations in~\cite{Kiesel11-1} and find that the quantum efficiency only leads to rescaled parameters $\w,\alpha$ of the witness,
\begin{equation}
	\eta{\rm Tr}\{\hat \rho_{\eta} \hat W_\w(\alpha)\} = {\rm Tr}\{\hat \rho \hat W_{\sqrt{\eta}\w}(\alpha/\sqrt{\eta})\}.
\end{equation}
Since it is known that the nonclassical effects become more pronounced with increasing $\w$, a lower quantum efficiency may make their significant verification more difficult. However, as a matter of principle, it does not change the property of a quantum state of being nonclassical or not. Therefore, for any positive quantum efficiency it is possible to verify nonclassicality if the number of recorded data points is sufficiently large.

\subsection{Statistical uncertainties}

Since the photon number probabilities $p_n(\alpha,\eta)$ are measured experimentally, they are subject to a statistical uncertainty, stemming from the finite number of measurements. Typically, they are calculated from the number of occurrences $N_n$ of a specific photon number $n$, divided by the total number of measurements $N$. Since the $N_n$ follow a multinomial distribution, with probabilities $p_n(\alpha,\eta)$ for each outcome, their covariance matrix is given by~\cite{StatisticsBook}
\begin{equation}
  \mathrm{C}(N_m,N_n) = N p_n(\alpha,\eta)(\delta_{mn}- p_m(\alpha,t)).
\end{equation}
Consequently, the covariance of the estimated probabilities equals to $\mathrm{C}(N_m,N_n)/N^2$. Now, the expectation value (\ref{eq:reconstruct:P}) is a linear function of the photon number probabilities. Therefore, its variance can be exactly estimated as
\begin{equation}
  {\rm Var}(\langle\hat W_{\w}(\alpha)\rangle_\eta) = \frac{1}{N^2}\sum_{m,n = 0}^\infty \mathrm{C}(N_m,N_n) \Omega_{\w,m}\Omega_{\w,n}.
\end{equation}
Obviously, the evaluation of statistical errors is straightforward and very simple. Moreover, we note that the  nonclassicality test is performed on the quantum-mechanical level of uncertainty, since only measurements of a single observable $\hat n(\alpha)$ are required~\cite{Kiesel-samplingnoise}.

\subsection{Systematic error estimation}

In practice, we cannot measure an infinite number of photon number probabilities $p_n$. However, we also know that all
states with a finite maximum photon number are nonclassical. Therefore, it is necessary to estimate the error which
arises due to the restriction of the state in Fock basis, in order to evaluate if a given state is nonclassical or the observed negativities are only due to the truncation of the Fock basis.

Let us assume that we can measure the probabilities of the Fock states $\ket 0$ to $\ket N$. Therefore, we estimate the
expectation value of the witness by 
\begin{equation}
  \langle\hat W_{\w}(\alpha)\rangle_\eta \approx \sum_{n=0}^N p_n(\alpha,\eta) \Omega_{\w,n}.
\end{equation}
The corresponding error therefore is given by
\begin{equation}
  \Delta W_{\w}(\alpha,\hat \rho) = \sum_{n=N+1}^\infty p_n(\alpha,\eta) \Omega_{\w,n}
\end{equation}
For a theoretically known state $\hat \rho$, it is not difficult to evaluate this quantity. However, we do not want to use assumptions about the state we examine. Therefore, we have to find a suitable alternative. Here, we ask for the maximum of the deviation $\Delta W_{\w}(\alpha,\hat \rho)$ for all classical states,
\begin{equation}
  \Delta W_{\w,\rm cl.}
    =\max_{\hat \rho_{\rm cl.}}\left[\Delta W_{\w}(\alpha,\hat \rho_{\rm cl.})\right].
  \label{eq:def:syst:err}
\end{equation}
If we can estimate this number, we know that if the state is classical, its expectation value of the nonclassicality witness is bounded from above by
\begin{equation}
  \langle\hat W_{\w}(\alpha)\rangle_{\rm real.} \leq \langle\hat W_{\w}(\alpha)\rangle_{\rm meas.} + \Delta W_{\w,\rm cl.}.
\end{equation}
If the right side of this equation still shows significant negativities, the state cannot be classical, but must be a nonclassical one.

As one can imagine, the calculation of the systematic error is more involved, such that we give the details in the Appendix. For practical reasons, we do not estimate the maximum with respect to all classical states, but restrict ourselves to states for which the probability for counting more than $N$ photons is less than some threshold $1-p_{\rm r}$, with $p_{\rm r}$ being the total probability of all resolved photon numbers.  This restriction is based on the assumption that the recorded data already gives sufficient information on the photon statistics, and the photon counts which cannot be resolved by the detector are negligible. It turns out that the systematic error is independent of the point $\alpha$, therefore we already omitted the argument in $\Delta W_{\w,\rm cl.}$. Finally, we show that the maximum can be found as the extremal point of a function of two real parameters, representing amplitudes of coherent states, cf.~Eqs.~\eqref{eq:optimization:objective:1}, \eqref{eq:optimization:objective:2}. This makes the calculation feasible for practical applications. 

\section{Example: single photon added thermal state}

Let us consider a single photon added thermal state (SPATS)~\cite{AgarwalTara} with a mean photon number $\bar n = 0.8$, and a preparation efficiency $\eta_{\rm prep.} = 0.5$. It is known that this state exhibits neither squeezing nor sub-Poissonian statistics, arising from Eq.~(\ref{eq_noncl_cond}) with $\hat W = :\!\!(\hat x - \langle\hat x\rangle)^2\!\!:$ and \mbox{$\hat W = :\!\!(\hat n - \langle\hat n\rangle)^2\!\!:$} respectively~\cite{Shchukin05}. Therefore, these two well-known nonclassicality witnesses are not able to verify nonclassicality of this particular state. 
It is also important that under the chosen conditions 
the  Wigner function is positive semidefinite~\cite{Bellini}, and the characteristic function does not show nonclassicality of first order~\cite{Ri-Vo}.
Therefore, it is a nontrivial task to witness the nonclassicality of the state under study.

We assume that the quantum efficiency of the detector equals to $\eta_D = 0.5$, leading to an overall efficiency of $\eta = 0.25$. Moreover, let us suppose that the detector may resolve the photon number statistics for up to $N = 15$ photons. The initial beam splitter has a transmissivity $t = 0.99$. We further assume that we perform %$100000$ 
$10^5$ measurements per coherent displacement $\alpha$. The probability of finding not more than $15$ photons equals to $p_{\rm r} = 0.995$ for coherent displacements with $|\alpha| < 3$. As a nonclassicality witness, we choose the example given in Eq.~\eqref{eq:triangular:filter}. Then, the matrix elements~\eqref{eq:matrix:elements} are given by
\begin{equation}
	\Omega_{\w,n} =  \frac{\w^2}{16} \sum_{m=0}^n \frac{(-\w^2/4)^m}{[(m+1)!]^2}\binom{2m+2}{m} \frac{m!}{(n-m)!}.
\end{equation}
Since SPATS are phase symmetric and similar to the photon, we may test the witness at the origin of phase space, i.e.~for $\alpha = 0$. 

\begin{figure}[h]
  \includegraphics[width=0.9\columnwidth]{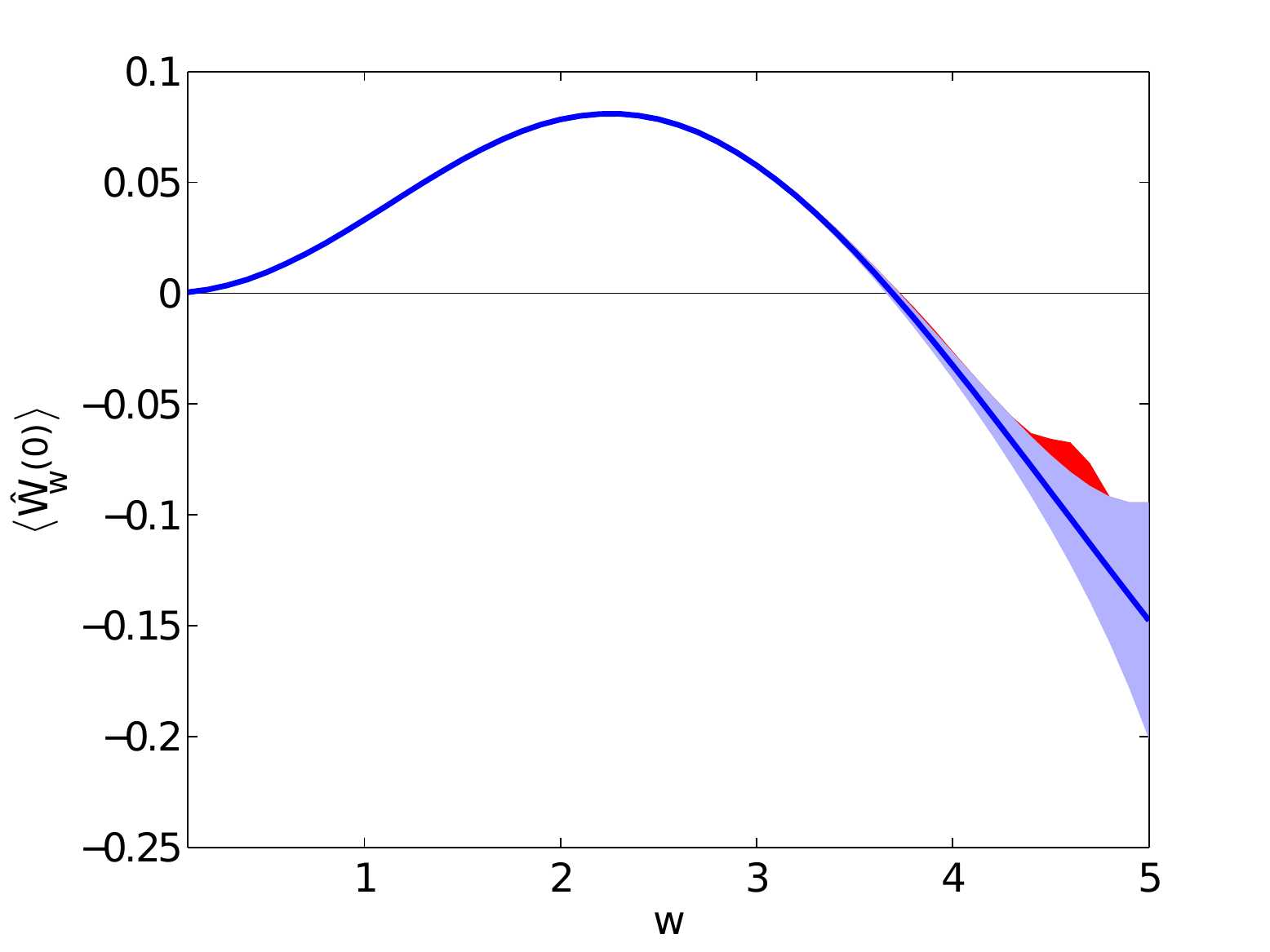}
  \caption{Dependence of the expectation value of the witness $\hat W_\w(0)$ on the filter width, for a SPATS with mean thermal photon number $\bar n = 0.8$ and overall efficiency $\eta = 0.25$. The blue shaded area corresponds to a statistical error of one standard deviation, the red shaded area shows the systematic error due to the truncation of the Fock space.}
  \label{fig:P:Omega:0}
\end{figure}
In Fig.~\ref{fig:P:Omega:0}, the expectation value of $\hat W_{\w}(0)$ is shown in dependence of $\w$ together with the statistical and systematic uncertainty. We note that the systematic error strongly depends on the value of $\w$. At some points between $\w = 4$ and $\w = 5$, it plays an important role, while it can be neglected for other values of $\w$. For sufficiently large $\w$, the expectation value becomes clearly negative.  The optimal significance is achieved for $\w = 4.2$. For this choice of the filter width our result is $\langle\hat W_\w(0)\rangle + \Delta W_{\w,\rm cl.} = -6.1\sqrt{{\rm Var}(\langle\hat W_{\w}(0)\rangle)}$, which is a significance of more than six standard deviations.

\begin{figure}[h]
  \includegraphics[width=0.9\columnwidth]{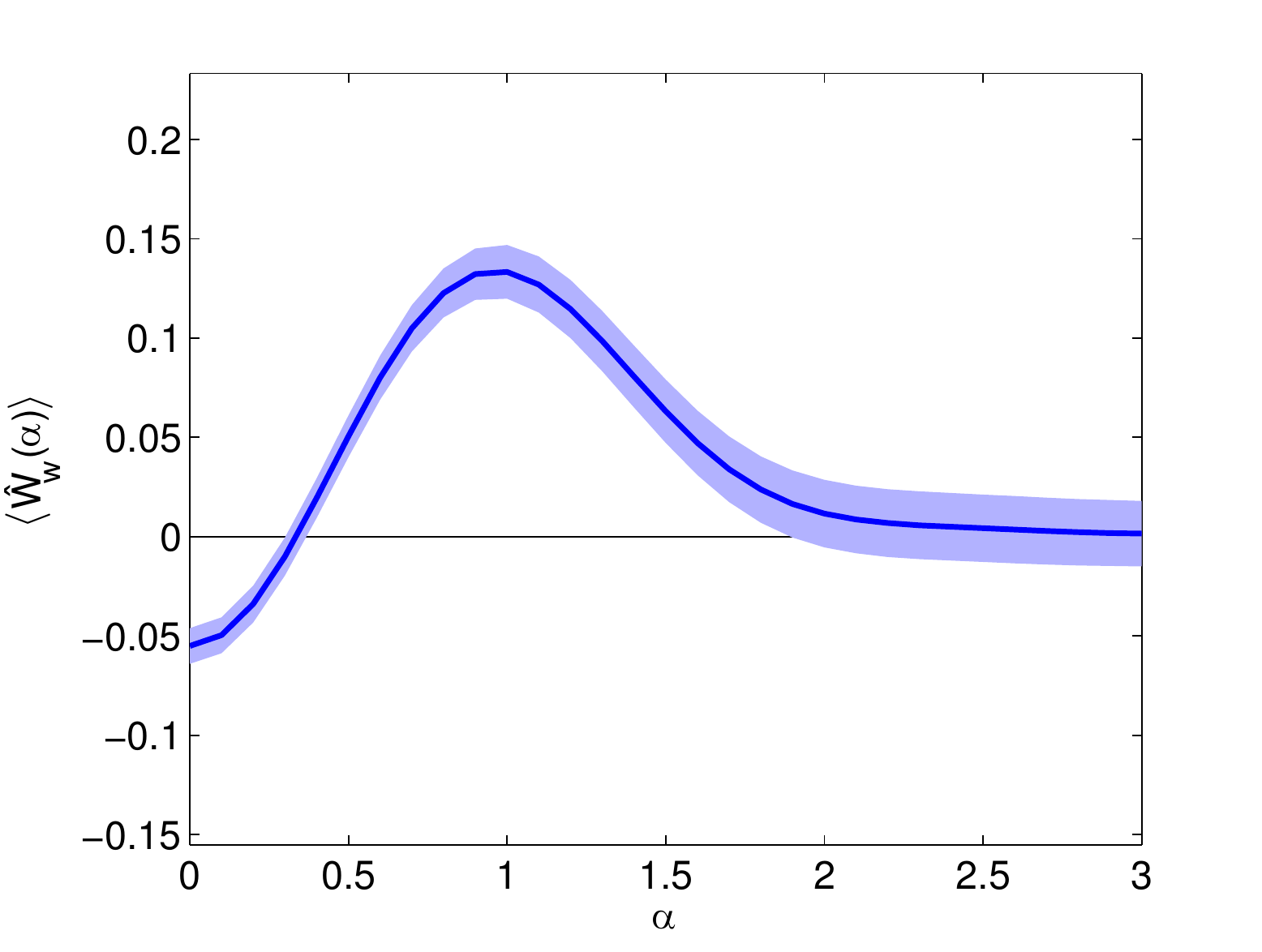}
  \caption{Dependence of the expectation value of the witness on the coherent displacement $\alpha$, for a SPATS with mean thermal photon number $\bar n = 0.8$ and overall efficiency $\eta = 0.25$,  and an optimal filter width $\w = 4.2$.  The blue shaded area corresponds to a statistical error of one standard deviation, the systematic error is negligible for this width parameter.}
  \label{fig:P:Omega}
\end{figure}

The dependence of the expectation values of $\hat W_{\w}(\alpha)$ on the coherent displacement $\alpha$ is shown in Fig.~\ref{fig:P:Omega}. The width $\w$ is chosen to be the value for optimal significance. The systematic error is negligible in this case. We observe the largest negativity for $\alpha = 0$. There, nonclassicality is verified with sufficiently high significance. We note that the curve also shows the course of the nonclassicality filtered $P$ function  of the SPATS, cf.~\cite{Kiesel11-1, Kiesel-witness}

Let us conclude this section with a brief discussion of the photon number resolution which plays a key role for the method under study. Whenever for the light source under study a properly mode-matched local oscillator (LO) is available,
the photon number resolution can easily realized by balanced homodyning with a phase randomized LO~\cite{Raymer}. For the considered example of a SPATS this method could be implemented~\cite{Bellini}, so that the required photon number resolution of up to $15$ photons is easily realized. For other radiation sources 
a proper LO may not be available. For such situations nowadays different realizations of photon number resolving detectors are under study~\cite{TMD1,TMD2,TMD3,Yamamoto,Bondani,TMD4}. The theory of such detectors shows that the measured statistics only slowly converges to that of a truly photon number resolving detector~\cite{SVA1}. However, one may directly use a measured sub-binomial statistics to certify nonclassicality~\cite{SVA2}. The further development of such direct methods for the general witnessing of nonclassicality is a demanding task for future research.

\section{Summary and Conclusions}

We have analyzed a simple experimental setup to measure the expectation value of an arbitrary nonclassicality witness. The coherent displacement $\alpha$ can be chosen by the experimental setup, only the width parameter $\w$ has to be introduced by a suitable post-processing procedure. Its optimal value depends on the quantum state under study and on the amount of available recorded data. It can be properly adjusted during the rather elementary computational data analysis. 

The applicability of our method has been simulated under realistic conditions for a single-photon added thermal state. For the analyzed quantum state,  the commonly used nonclassicality criteria fail to identify its nonclassicality. 
The state is neither sub-Poissonian nor quadrature squeezed, the Wigner function is positive semidefinite, and the characteristic function does not display nonclassicality directly by exceeding the boundaries set by its vacuum noise level.
Even under such complicated conditions and after taking systematic and statistical uncertainties into account, our witness approach clearly verifies the nonclassicality with a good significance.  

\section*{Acknowledgments} One of the authors (T.K.) gratefully acknowledges A.~Kiesel for many useful hints on linear optimization. This work was supported by the Deutsche Forschungsgemeinschaft through SFB 652.

\appendix

\section*{Appendix: Systematic error estimation}\label{app:syserr}

Here, we show how to calculate the systematic error 
\begin{equation}
  \Delta W_{\w,\rm cl.}
    =\max_{\hat \rho_{\rm cl.}}\left[\sum_{n=N+1}^\infty p_n(\alpha, \eta)\Omega_{\w,n}\right].
\end{equation}
All classical states can be commonly described by a nonnegative $P$ function, therefore we take the optimization with respect to all $P$ functions admitting an interpretation as a classical probability density. Clearly, the objective function is linear in $P(\alpha)$. Therefore, we have to solve a generalized linear optimization problem, with constraints
\begin{equation}
  P(\alpha) \geq 0,\qquad \int P(\alpha)d^2\alpha = 1.
\end{equation}

We solve this task in several steps. First, we formulate the optimization problem and transfer it into a more simple form. Second, we discretize the problem and apply standard knowledge about linear optimization problems to uncover the structure of the solution. Last, we argue why this structure also holds in the continuous case, and present a simple equivalent constrained optimization problem, which only requires to maximize with respect to two real variables.

\subsection{Optimization problem}
Let us examine the objective function for coherent states $\hat \rho_{\rm cl.} = \ket\beta\bra\beta$:
\begin{equation}
  \Delta W_{\w}(\alpha,\ket\beta,\eta) = \!\sum_{n=N+1}^\infty\! \frac{|\beta - \alpha|^{2n}\eta^{n}}{n!} e^{-|\beta - \alpha|^2 \eta}\Omega_{\w,n}.\label{eq:deviation:coherent:state}
\end{equation}
Here we already inserted the photon number distribution $p_n(\alpha,\eta)$. Obviously, the objective function only depends on the value of $|\alpha - \beta| \sqrt{\eta}$. Therefore, we may write
\begin{equation}
  \Delta W_{\w}(\alpha,\ket\beta,\eta) = \Delta W_{\w}(|\alpha-\beta| \sqrt{\eta},\ket 0,1).\label{eq:Delta:W:depends:on:absval}
\end{equation}
For arbitrary states, characterized by a $P$ function $P(\beta)$, the systematic error can be obtained via
\begin{eqnarray}
 \hspace{-1.5em}\Delta W_{\w}(\alpha,\hat\rho,\eta) &=& \int d^2\beta P(\beta) \Delta W_{\w}(\alpha,\ket\beta,\eta)\nonumber\\
	&=& \!\! \int d^2\beta P(\beta) \Delta W_{\w}(|\alpha-\beta| \sqrt{\eta}, \ket 0,1).\label{eq:objective:general:P}
\end{eqnarray}
A simple substitution $\gamma = (\alpha-\beta) \sqrt{\eta}$ brings us to
\begin{equation}
 \Delta W_{\w}(\alpha,\hat\rho,\sqrt{\eta}) 
	= \int d^2\gamma P(\alpha-\gamma/\sqrt{\eta}) \Delta W_{\w}(|\gamma|, \ket 0,1).
\end{equation}
Finally, we introduce a scaled radial marginal of the $P$ function,
\begin{equation}
	R_\alpha(b) = \frac{b}{\eta} \int_0^{2\pi} P(\alpha-b e^{i\varphi}/\sqrt{\eta}) d\varphi,
\end{equation}
which is defined for nonnegative real arguments $b$. For classical states, this quantity fulfills all requirements of a probability density. The objective function~(\ref{eq:objective:general:P}) is then expressed by
\begin{equation}
 \Delta W_{\w}(\alpha,\hat\rho,\eta) = \int_0^\infty  R_\alpha(b) \Delta W_{\w}(b,\ket 0,1 ) db.\label{eq:objective:general:R} 
\end{equation}
Until now, maximization $\Delta W_{\w}(\alpha,\hat\rho,\eta)$ over all classical states $\hat\rho_{\rm cl.}$ is equivalent to maximization over all probability densities $R_\alpha(b)$. 

We would like to add one additional constraint on $R(b)$. Let us assume that the probability for finding more than $N$ photons is less than some constant, i.e.
\begin{equation}
  \sum_{n=N+1}^\infty p_n(\alpha,t) = 1 - \sum_{n=0}^N p_n(\alpha,t) \leq  1 - p_{\rm r}.
\end{equation}
This accounts for the fact that we did not observe more than $N$ photons in an experiment, therefore the probability for doing so should be quite small. We can express this condition in terms of the $P$ function of the input state as
\begin{equation}
   \int d^2\beta P(\beta) \sum_{n=0}^N \frac{|\beta - \alpha|^{2n}\eta^{n}}{n!} e^{-|\beta - \alpha|^2 \eta} \geq p_{\rm r}.
\end{equation}
By applying the same considerations as before, and introducing the abbreviation
\begin{equation}
 G(b) = \sum_{n=0}^N \frac{b^{2n}}{n!} e^{-b^2},
\end{equation}
we can express this inequality in terms of the density $R_\alpha(b)$ and obtain
\begin{equation}
    \int_0^\infty R_\alpha(b) G(b) db   \geq p_{\rm r}. \label{eq:constraint:R}
\end{equation}

In conclusion, we want to optimize the objective function Eq.~(\ref{eq:objective:general:R}) with respect to the probability density $R(b)$, satisfying the natural constraints 
\begin{equation}
	\int R_\alpha(b) db = 1,\qquad R_\alpha	(b) \geq 0,\label{eq:opt:problem:prob:constraint}
\end{equation}
as well as our additional requirement Eq.~(\ref{eq:constraint:R}). This is a linear optimization problem on the convex set of admissible $R_\alpha(b)$. Since the objective function as well as the constraints do not depend on $\alpha$, the solution does not as well, and we will omit the index $\alpha$ from now on.

Before we continue with the solution of this problem, let us comment on two properties of the function $G(b)$. On the one hand, it can be easily seen that $G(b)$ is strictly increasing with $N$ for all fixed $b > 0$. On the other hand, it is decreasing with increasing $b$ for all $N > 0$. The latter fact becomes clear when one calculates the derivative,
\begin{equation}
  G'(b) = - 2 \frac{b^{2 N + 1}}{N!} e^{-b^2},
\end{equation}
which is strictly negative.

\subsection{Structure of the solution}

It is well-known that the maximum of a linear optimization problem is achieved at vertices of the admissible set of $R(b)$~\cite{LinearOptimization}. For this purpose, we have to examine the structure of the set of functions satisfying the constraints (\ref{eq:constraint:R}) and (\ref{eq:opt:problem:prob:constraint}). As a first step, we analyze a discrete analogue of this set. Afterwards, we will apply the results to the continuous case.

For discretisation, we replace the integrals in (\ref{eq:constraint:R}) and (\ref{eq:opt:problem:prob:constraint}) by Riemannian sums with $K+1$ interpolation points, i.e.
\begin{equation}
	\int_0^\infty f(b) db \to \sum_{k=0}^K f(\Delta b\,k) \Delta b.
\end{equation}
By introducing discrete variables
\begin{equation}
  r_k = R(\Delta b\,k)\Delta b,\quad  g_k = G(\Delta b\,k)
\end{equation}
with indices $0 \leq k \leq K$, the constraints are replaced by
\begin{eqnarray}
	\sum_{k=0}^K r_k g_k &\geq& p_{\rm r}, \label{eq:opt:problem:constraint:1}\\
    \sum_{k=0}^K r_k &=& 1,\qquad r_k \geq 0.\label{eq:opt:problem:constraint:2}
\end{eqnarray}
The examination of such convex sets of $(r_k)_{k=0}^K$ is a standard problem. The vertices  can be determined as follows: First, we rewrite the inequality~(\ref{eq:opt:problem:constraint:1}) as an equality with an additional constrained slack variable y, 
\begin{equation}
   \sum_{k=0}^K r_k g_k - y = p_{\rm r},\qquad y \geq 0.
\end{equation}
Then, the set of constraints can be formulated in the form
\begin{eqnarray}
  \left(
  \begin{array}{c c c | c}
      1	& \ldots & 1 & 0\\
      g_0 & \ldots & g_K & -1
  \end{array}\right) 
  \left(
  \begin{array} {c} \vec r\\ y \end{array}\right) 
  &=& \left(\begin{array} {c} 1\\ p_{\rm r} \end{array}\right) \\
  r_k \geq 0,\qquad y \geq 0
\end{eqnarray}
To find a vertex, we set an arbitrary selection of $K$ components of the vector $(\vec r,y)^{T}$ to zero, and solve
the equation for the remaining two components. If the resulting solution has only nonnegative entries, it is a vertex of the polyhedron of feasible solutions of the optimization problem. We have to distinguish two cases:
\begin{enumerate}
 \item We set $K$ elements of the vector $\vec r$ to zero. Therefore, we are left with some element $r_k$ and the
slack variable $y$, and consider the equation
    \begin{equation}
      \left(
      \begin{array}{c c}
	1	& 0\\
	g_k 	 & -1
      \end{array}\right) 
      \left(
    \begin{array} {c} r_k\\ y \end{array}\right) 
    = \left(\begin{array} {c} 1\\ p_{\rm r} \end{array}\right)
    \end{equation}
    The solution of this equations reads as
    \begin{equation}
      r_k = 1,\quad y_k = g_k - p_{\rm r}.
    \end{equation}
    Nonnegativity of $y$ has to be taken into consideration, therefore $g_k= G(\Delta b\,k)$ should be larger than $p_{\rm r}$. Since this sequence of numbers is strictly decreasing with increasing $k$, the index $k$ has to be sufficiently small.
  \item We set $y= 0$ and $K-1$ components of the vector $\vec r$ to zero. Hence, we are left with two components $r_k$ and $r_l$, $k < l$. This leads to 
      \begin{equation}
      \left(
      \begin{array}{c c}
	1	& 1\\
	g_k 	 & g_l
      \end{array}\right) 
      \left(
    \begin{array} {c} r_k\\ r_l \end{array}\right) 
    = \left(\begin{array} {c} 1\\ p_{\rm r} \end{array}\right),
    \end{equation}
   and its solution is given by
    \begin{equation}
	r_k = \frac{p_{\rm r} - g_l}{g_k-g_l}, \qquad r_l = \frac{g_k-p_{\rm r}}{g_k-g_l}.
    \end{equation}
  Since the sequence $g_k = G(\Delta b\,k)$ is decreasing, the denominator is strictly positive for pairs $k < l$. Therefore, we have to require that $l$ is sufficiently large such that $p_{\rm r} > g_l$, and $k$ is sufficiently small such that $g_k \geq p_{\rm r}$. 
\end{enumerate}
In conclusion, we have found that only two different types of vertices exist: The one only have a single element of the sequence $r_k$ being different from zero, provided that $k$ is sufficiently small. The others have two elements $r_k, r_l$ being different from zero. We emphasize that this is independent of the number of sampling points $K$. Therefore, we may improve the accuracy of the approximation by an Riemannian sum by increasing the number of sampling points and decreasing their distance $\Delta b$. 

\subsection{Solution of the continuous optimization problem}

Knowing the vertices of the discretized admissible set, let us return to the original problem. We do this by taking the limit $\Delta b \to 0$, $K \to \infty$. The approximation of an integral by a Riemannian sum effectively replaces the function $R(b)$ by a step function 
\begin{equation}
  R_{\Delta b}(b) = \sum_{k=0}^K r_k {\rm rect}_{\Delta b}(b - \Delta b k),
\end{equation}
with ${\rm rect}_{\Delta b}(x) = 1/\Delta b$ for all $0 \leq x < \Delta b$ and ${\rm rect}_{\Delta b}(x) = 0$ elsewhere. All rectangles have unit area, satisfying 
\begin{equation}
 \int _{-\infty}^\infty {\rm rect}_{\Delta b}(x) = 1.
\end{equation}
for all positive $\Delta b$. The vertices with a single nonzero component then read as
\begin{equation}
	R^{(1)}_{v,\Delta b} (b) = {\rm rect}_{\Delta b}(b - \Delta b\, k). 
\end{equation}
In the limit of $\Delta b \to 0$, the function ${\rm rect}_{\Delta b}(x)$ converges to the $\delta$-distribution, centered at some point $b_1$. Therefore, the vertex takes the form
\begin{equation}
	R^{(1)}_{b_1} (b) = \delta(b-b_1).\label{eq:opt:solution:1}
\end{equation}
The constraint $g_k \geq p_{\rm r}$ has to be replaced by $G(b_1) \geq p_{\rm r}$.  
Analogously, the vertices with two nonzero components $r_k,r_l$ read as
\begin{eqnarray}
  R^{(2)}_{v,\Delta b} (b) &=& \frac{p_{\rm r} - g_l}{g_k-g_l} {\rm rect}_{\Delta b}(b - \Delta b\, k) \nonumber\\
		&& + \frac{g_k-p_{\rm r}}{g_k-g_l}{\rm rect}_{\Delta b}(b - \Delta b\, l).
\end{eqnarray}
In the continuous limit, we have to replace the rectangular functions by $\delta$-functions and the coefficients $g_{k} = G(\Delta b\,k)$ by $G(b_{1,2})$. Hence, these vertices are given by
\begin{eqnarray}
  R^{(2)}_{b_1,b_2}(b) &=& \frac{p_{\rm r} - G(b_2)}{G(b_1)-G(b_2)}\delta( b-b_1) \nonumber\\
		&&+ \frac{G(b_1)-p_{\rm r}}{G(b_1)-G(b_2)} \delta( b-b_2),\label{eq:opt:solution:2}
\end{eqnarray}
where the parameter $b_1, b_2$ satisfy 
\begin{equation}
	G(b_2) < p_{\rm r} \leq G(b_1). \label{eq:opt:constraint:b1:b2}
\end{equation}

In conclusion, we have demonstrated that the objective function (\ref{eq:objective:general:R}) is maximized for some function $R(b)$ having the form (\ref{eq:opt:solution:1}) or (\ref{eq:opt:solution:2}). This solution only depends on one real parameter $b_1$ or two real parameters $b_1,b_2$ respectively. Therefore, in order to find the systematic error, one has to find the global maxima of the two objective functions
\begin{eqnarray}
	\Delta W^{(1)}_{\w}(b_1) &=& t^2 \Delta W_{\w}(b_1,\ket 0,1 )\label{eq:optimization:objective:1}\\
	\Delta W^{(2)}_{\w}(b_1,b_2) &=& t^2 \left[\frac{p_{\rm r} - G(b_2)}{G(b_1)-G(b_2)}\Delta W_{\w}(b_1,\ket 0,1 ) \right.\nonumber\\
			+&&\!\!\!\!\!\left.\frac{G(b_1)-p_{\rm r}}{G(b_1)-G(b_2)} \Delta W_{\w}(b_2,\ket 0,1 )\right],\label{eq:optimization:objective:2}
\end{eqnarray}
with $G(b_2) < p_{\rm r} \leq G(b_1)$. The largest solution gives the systematic error $\Delta W_{\w,\rm cl.}$.


\begin{thebibliography}{99}
\bibitem{Schrodinger} E.~Schr\"odinger, Naturwiss. {\bf 14}, 664 (1926).
\bibitem{Sudarshan} E.~C.~G.~Sudarshan, Phys. Rev. Lett. {\bf 10}, 277 (1963).
\bibitem{Glauber} R.~J.~Glauber, Phys. Rev. {\bf 131}, 2766 (1963).
\bibitem{TitulaerGlauber} U.~M.~Titulaer and R.~J.~Glauber, Phys. Rev. {\bf 140}, B676 (1965).
\bibitem{Wineland} D. Leibfried, D. M. Meekhof, B. E. King, C. Monroe, W. M. Itano, and D. J. Wineland, Phys. Rev. Lett. {\bf 77}, 4281 (1996).
\bibitem{Kenfack} A. Kenfack and K. \.Zyczkowsky, J. Opt. B {\bf 6}, 396 (2004).
\bibitem{Kiesel10} T. Kiesel and W. Vogel, Phys. Rev. A {\bf 82}, 032107 (2010).
\bibitem{Kiesel11-1} T. Kiesel, W. Vogel, M. Bellini, and A. Zavatta, Phys. Rev. A {\bf 83}, 032116 (2011).
\bibitem{Kiesel11-2} T. Kiesel, W. Vogel, B. Hage, and R. Schnabel, Phys. Rev. Lett. {\bf 107}, 113604 (2011).
\bibitem{Kiesel12} T. Kiesel, W. Vogel, S. L. Christensen, J.-B. B\'eguin, J. Appel, and E. S. Polzik, arXiV:quantph/1207.3314 (2012).
\bibitem{Shchukin05} E. Shchukin, T. Richter, and W. Vogel, Phys. Rev. A {\bf 71}, 011802(R) (2005).
\bibitem{Korbicz05} J. K. Korbicz, J. I. Cirac, J. Wehr, and M. Lewenstein, Phys. Rev. Lett. {\bf 94}, 153601 (2005).
\bibitem{Shchukin05b} E. V. Shchukin and W. Vogel, Phys. Rev. A {\bf 72}, 043808 (2005).
\bibitem{Luis79}  \'A. Rivas and A. Luis, Phys. Rev. A {\bf 79}, 042105, (2009).
\bibitem{Kiesel-witness} T. Kiesel and W. Vogel, Phys. Rev. A {\bf 85}, 062106 (2012).
\bibitem{Wallentowitz96} S. Wallentowitz and W. Vogel, Phys. Rev. A {\bf 53}, 4528, (1996).
\bibitem{Kiesel-samplingnoise} T. Kiesel, Phys. Rev. A {\bf 85}, 052114 (2012).
\bibitem{NecSufPhotonNumber} R. Simon, M. Selvadoray, Arvind, and N. Mukunda, arXiV:quantph/9709030 (1997).
\bibitem{StatisticsBook} M. Evans, N. Hastings, and B. Peacock, \emph{Statistical Distributions}, pp. 134–136, Wiley New York (2000).
\bibitem{AgarwalTara} G. S. Agarwal and K. Tara, Phys. Rev. A {\bf 46}, 485 (1992).
\bibitem{Bellini} A. Zavatta, V. Parigi, and M. Bellini, Phys. Rev. A {\bf 75}, 052106 (2007).
\bibitem{Ri-Vo} W. Vogel, Phys. Rev. Lett. {\bf 84}, 1849 (2000); T. Richter and W. Vogel, Phys. Rev. Lett. {\bf 89}, 283601 (2002).
\bibitem{Raymer} M. Munroe, D. Boggavarapu, M. E. Anderson, and M. G. Raymer, Phys. Rev. A {\bf 52}, R924 (1995).
\bibitem{TMD1} J. Rehacek, Z. Hradil, O. Haderka, J. Perina, Jr., and M. Hamar, Phys. Rev. A {\bf 67}, 061801(R) (2003).
\bibitem{TMD2} D. Achilles, Ch. Silberhorn, Cezary \'Sliwa, K. Banaszek, and I. A. Walmsley, Opt. Lett. {\bf 28}, 2387 (2003).
\bibitem{TMD3} M. J. Fitch, B. C. Jacobs, T. B. Pittman, and J. D. Franson, Phys. Rev. A {\bf 68}, 043814 (2003).	
\bibitem{Yamamoto} E. Waks, E. Diamanti, B. C. Sanders, S. D. Bartlett, and Y. Yamamoto, Phys. Rev. Lett. {\bf 92}, 113602 (2004).
\bibitem{Bondani} M. Bondani, A. Allevi, A. Agliati, and Andreoni, J. Mod. Opt. 
{\bf 56}, 226 (2009).	
\bibitem{TMD4} G. Brida, M. Genovese, M. Gramegna, A. Meda, F. Piacentini, P. Traina, E. Predazzi, S. Olivares, and M. G. A. Paris, Adv. Sci. Lett. {\bf 4}, 1 (2011).
\bibitem{SVA1} J. Sperling, W. Vogel, and G. S. Agarwal,
       Phys. Rev. A {\bf 85}, 023820 (2012).
\bibitem{SVA2} J. Sperling, W. Vogel, and G. S. Agarwal,
      Phys. Rev. Lett., in press; arXiv:1205.4669.
\bibitem{LinearOptimization} A. Schrijver, \emph{Theory of linear and integer programming}, Wiley Chichester (1998).

\end{thebibliography}
\end{document}